\documentclass[aps,preprintnumbers,floatfix,article,amsmath,amssymb,floatfix,10pt,prd,superscriptaddress,nofootinbib,natbib]{revtex4-2}
\usepackage{bm}
\usepackage{amsfonts}
\usepackage{latexsym}
\usepackage[latin1]{inputenc}
\usepackage{graphicx}
\usepackage{amsmath}
\usepackage{palatino}
\usepackage{mathpazo}
\usepackage{textcomp}
\usepackage{float}
\usepackage{booktabs}
\usepackage{dcolumn}
\usepackage{ragged2e}
\usepackage{amsmath}
\usepackage{xcolor}
\usepackage{orcidlink}
\usepackage{epsfig}
\usepackage{caption}
\usepackage{subcaption}
\usepackage{commath}
\usepackage{hyperref}
\hypersetup{colorlinks,citecolor=purple}
\hypersetup{colorlinks=true,linkcolor=red,filecolor=magenta,    urlcolor=blue}

\def\jnl@style{\it}
\def\aaref@jnl#1{{\jnl@style#1}}

\def\aaref@jnl#1{{\jnl@style#1}}

\def\aj{\aaref@jnl{AJ}}                   
\def\apj{\aaref@jnl{ApJ}}                 
\def\apjl{\aaref@jnl{ApJ}}                
\def\apjs{\aaref@jnl{ApJS}}               
\def\apss{\aaref@jnl{Ap\&SS}}             
\def\aap{\aaref@jnl{A\&A}}                
\def\aapr{\aaref@jnl{A\&A~Rev.}}          
\def\aaps{\aaref@jnl{A\&AS}}              
\def\mnras{\aaref@jnl{Mon.~Not.~Roy.~Astron.~Soc.}}             
\def\prd{\aaref@jnl{Phys.~Rev.~D}}        
\def\prc{\aaref@jnl{Phys.~Rev.~C}}  
\def\prl{\aaref@jnl{Phys.~Rev.~Lett.}}    
\def\qjras{\aaref@jnl{QJRAS}}             
\def\skytel{\aaref@jnl{S\&T}}             
\def\ssr{\aaref@jnl{Space~Sci.~Rev.}}     
\def\zap{\aaref@jnl{ZAp}}                 
\def\nat{\aaref@jnl{Nature}}              
\def\aplett{\aaref@jnl{Astrophys.~Lett.}} 
\def\apspr{\aaref@jnl{Astrophys.~Space~Phys.~Res.}} 
\def\physrep{\aaref@jnl{Phys.~Rep.}}      
\def\physscr{\aaref@jnl{Phys.~Scr}}       
\def\commat{\aaref@jnl{Comm.~Math.~Phys.}}              
\def\science{\aaref@jnl{Science}}               
\def\cqg{\aaref@jnl{Classical Quant.~Grav.}}            
\def\jpcs{\aaref@jnl{JPCS}}                                     
\def\ijmpd{\aaref@jnl{Int.~J.~Mod.~Phys.~D}}                    
\def\grg{\aaref@jnl{Gen.~Relat.~Gravit.}}               
\def\rpp{\aaref@jnl{Rep.~Prog.~Phys.}}          
\def\npa{\aaref@jnl{Nucl.~Phys.~A}}        
\def\lrr{\aaref@jnl{Living Rev.~Rel.}}                   
\def\jcap{\aaref@jnl{J.~Cosmology Astropart.~Phys.}}    
\def\rmp{\aaref@jnl{Rev.~Mod.~Phys.}}   
\def\epjc{\aaref@jnl{Eur.~Phys.~J.~C}} 
\def\plb{\aaref@jnl{~Phy.~Lett.~B}} 
\def\mpla{\aaref@jnl{Mod.~Phy.~Lett.~A}} 
\def\arxiv{\aaref@jnl{arxiv.org}}

\allowdisplaybreaks[1]

\begin{document}

\color{black}       

\title{Accelerating Cosmological Model with Scalar Field in $f(R,\mathcal{L}_{m})$ Gravity}
\author{Y. Kalpana Devi\orcidlink{0009-0001-2686-7281}}
\email{kalpanayengkhom123@gmail.com}
\affiliation{Department of Mathematics, Birla Institute of Technology and Science, Pilani, Hyderabad Campus, Jawahar Nagar, Kapra Mandal, Medchal District, Telangana-500078, India.}

\author{Rahul Bhagat\orcidlink{0009-0001-9783-9317}}
\email{rahulbhagat0994@gmail.com}
\affiliation{Department of Mathematics, Birla Institute of Technology and Science, Pilani, Hyderabad Campus, Jawahar Nagar, Kapra Mandal, Medchal District, Telangana-500078, India.}

\author{B. Mishra\orcidlink{0000-0001-5527-3565}}
\email{bivu@hyderabad.bits-pilani.ac.in}
\affiliation{Department of Mathematics, Birla Institute of Technology and Science, Pilani, Hyderabad Campus, Jawahar Nagar, Kapra Mandal, Medchal District, Telangana-500078, India.}

\date{\today}

\begin{abstract}

In this work, we investigate the cosmological dynamics of $f(R,\mathcal{L}_m)$ gravity using two complementary scenarios: without a scalar field and with a minimally coupled generalized scalar field. For the case without a scalar field, we consider a functional form with linear and exponential dependence on the matter Lagrangian and perform a dynamical system analysis. The resulting autonomous system admits a matter-dominated saddle configuration $C_1$ with $q=\frac{1}{2}$ and $\omega=0$, and a de Sitter attractor $C_2$ characterized by $q=-1$ and $\omega_{\rm eff}\to -1$. Due to the non-hyperbolic nature of the critical curves, Center Manifold Theory (CMT) is employed to establish the local asymptotic stability of the de Sitter solution. The cosmological evolution exhibits a transition from decelerated to accelerated expansion, with a transition redshift $z_t\approx0.7$ and a present value $q_0\approx-0.65$, in agreement with observations. We then extend the framework by including a minimally coupled generalized scalar field with an exponential self-interacting potential $V(\phi)=V_0e^{-n\phi}$. The extended autonomous system also contains a matter-dominated saddle point and a stable dark-energy-dominated attractor corresponding to a late-time de Sitter phase. The stability of the attractor is confirmed through CMT, and the evolution of the cosmological parameters demonstrates a smooth transition from a matter-dominated decelerating era to accelerated expansion with $\omega_{\rm eff}\to -1$. Also, in the absence of the scalar field, the matter-dominated configuration associated with a vanishing nonlinear contribution remains stable, whereas the inclusion of the scalar field transforms the matter era into a saddle configuration, thereby enabling the Universe to naturally evolve toward the late-time accelerated attractor. In both scenarios, the exponential power term equal to $-1$ corresponds to a late-time de Sitter phase. These results show that both scenarios lead to late-time cosmic acceleration, while the inclusion of the scalar field provides an additional dynamical mechanism for realizing a stable dark-energy-dominated Universe without invoking a cosmological constant.
 
\end{abstract}

\maketitle

\section{Introduction} \label{Introduction}

The scalar fields in cosmology are some hypothetical fields and are believed to play a crucial role in the understanding of late-time evolution of the Universe~\cite{Szyd_owski_2014}. The cosmological models with scalar fields are used to describe the early Universe phenomena such as inflation and the present dark energy phase~\cite{Guth_1981_23_347}. Since the scalar fields evolve in time, it may produce some dynamic explanation for dark energy~\cite{Roy_2022_36_101037}. In the quintessence phase, the scalar fields may show the accelerated expansion of the Universe from present to late time~\cite{Runkla_2018_98_084034,Smith_2020_160_267}. In addition, the scalar field is also considered a potential candidate for dark matter and arises naturally in string theory~\cite{Ortin_2004}. The compactification to four-dimensional spacetime from higher-dimensional spacetime can be realized~\cite{Saridakis_2021}. Since late-time cosmic acceleration behavior and the prediction of dark energy have changed the understanding of the Universe, modified theories of gravity have been introduced to address this prediction~\cite{B_hmer_2022}. One can consider any of the three approaches, such as curvature, torsion and non-metricity to frame the modified theories of gravity~\cite{Jimenez_2019_5_173,Sultana500354,Bhgat_2026_Adf,BHAGAT_2025_new}. The first modified gravity with curvature is the $f(R)$ gravity \cite{Carroll_2005_71_063513}, where $R$ is the Ricci scalar. When the functional dependence on $R$ introduces a scalar degree of freedom with an effective potential, then $f(R)$ gravity can be reformulated to an equivalent scalar-tensor theory \cite{Carroll_2004}. In $f(R, \mathcal{L}_m)$ gravity, a more generalized coupling has been introduced with the explicit dependence on the matter Lagrangian $\mathcal{L}_m$ \cite{Bohmer_2007}. The non-minimal coupling provides the deviation from geodesic motion for test particles, which results in an extra force orthogonal to the four-velocity \cite{Harko_2008}. The modified Friedmann equations in this framework contain additional terms arising from the coupling function, which can drive accelerated expansion at low redshifts while naturally recovering standard matter-dominated behavior at early times without the need to introduce an explicit cosmological constant $\Lambda$. The detailed formulation of $f(R, \mathcal{L}_m)$ gravity in Ref \cite{Harko_2010b}. The interaction between curvature and matter can significantly modify the evolution of the Universe by altering the Hubble expansion rate, affecting the growth of cosmic structures, and influencing other cosmological processes. From a scalar-field perspective, the $f(R,\mathcal{L}_m)$ cosmological framework enriches the dynamics by introducing scalar-like degrees of freedom sourced by both curvature and matter. Moreover, the nonminimal geometry--matter coupling uniquely determines the matter Lagrangian and leads to an extra force acting on matter, resulting in departures from standard geodesic motion even for dust fluids \cite{Harko_2010}. In particular, cosmological studies have shown that the geometry matter coupling can naturally account for late-time cosmic acceleration and may even alleviate the Big-Bang singularity without the need for a cosmological constant \cite{Gonccalves_2023}. For further discussions on the cosmological and astrophysical implications of $f(R,\mathcal{L}_m)$ gravity, see Refs.\cite{Harko_2010,Harko_2010a,BHAGAT2026100483,Savvas_2009,Valerio_2007,Valerio_2009,Gonccalves_2023}.\\

 In $f(R,\mathcal{L}_m)$ gravity, various studies have explored the implications of non-minimal coupling between curvature and matter for both theoretical consistency and observational viability. Wang et al. \cite{Wang_2012} applied the standard energy conditions to a specific $f(R, \mathcal{L}_m)$ model and derived the constraints on the parameters using cosmological observations. Faraoni \cite{Faraoni_2009} revisited the correct Lagrangian description of perfect fluid, $\mathcal{L}_1 = p$ versus $\mathcal{L}_2 = -\rho$. It has been shown that these formulations are equivalent under minimal coupling, but for non-minimally coupling, it leads to distinct physical predictions. Thakur et al.\cite{THAKUR2011309} examined the consequences of non-minimal gravitational couplings adopting specific functional forms of the coupling and comparing the behavior with the minimal coupling case. In another notable contribution, Nojiri et al. \cite{NOJIRI2004137} proposed that the accelerated expansion of the Universe could naturally arise from a decrease in curvature, with dark energy eventually dominating over standard matter. More recently, Ref. \cite{DEVI2024101640} presented an accelerating cosmological model in $f(R, \mathcal{L}_m)$ gravity, with its parameters tightly constrained using a combination of cosmological datasets and Gaussian reconstruction~\cite{devi2025late}, further strengthening the observational viability of such theories. Myrzakulova et al. \cite{Myrzakulova_2024} investigated the dark energy phenomenon within the framework of $f(R,\mathcal{L}_m)$ gravity by considering specific nonlinear models and demonstrated a transition of the deceleration parameter from an early decelerating phase to a late-time accelerating epoch. Also, non-linear $f(R,\mathcal{L}_m)$ model through a Bayesian analysis based on a parametrized effective equation-of-state parameter has been examined\cite{Myrzakulov_2023}. Koussour et al. \cite{Koussour2_2023} have analyzed a scalar-field dark energy model with a parametrized equation of state and have shown that a quintessence-like behavior capable of driving the present accelerated expansion of the Universe. Recent investigations have explored various astrophysical and cosmological implications of curvature-matter coupling theories. Gravitational wave propagation in such frameworks was analyzed by Bertolami et al. \cite{Bertolami_2018}, which shows good agreement with the most recent data. Dynamical system approaches were developed by Azizi and Yaraie \cite{Azizi:2014qsa} to characterize cosmological fixed points and their stability. On astrophysical scales, Tangphati et al. \cite{Tangphati_2025} and Priyobarta et al. \cite{PRIYOBARTA2026162} studied compact star configurations in $f(R,\mathcal{L}_m,T)$ gravity, demonstrating significant modifications in mass-radius relations and stability conditions. Furthermore, Bhardwaj and Ray \cite{BHARDWAJ2025101930} examined cosmological models within $f(R,\mathcal{L}_m)$ gravity, showing that the curvature-matter coupling can account for late-time acceleration. \\

Expanding on the pursuit of explaining cosmic acceleration beyond the traditional cosmological constant, a particularly intriguing proposition links the enigmatic, smoothly distributed energy component to a scalar field $\phi$. This evolving form of dark energy often referred to as dynamical $\Lambda$ or quintessence has attracted significant attention \cite{PhysRevD.62.023504,A2023169392,PhysRevD.61.023518,Bhagat_ASPdyna2024,BHAGAT2025101913}. Within the realm of curvature-based modifications to general relativity, $f(R)$ gravity has emerged as a leading candidate. A logical progression involves integrating a scalar field $\phi$ into this framework, leading to $f(R, \phi)$ gravity \cite{Farajollahi_2011}. This can be further generalized to $f(R, \phi, X)$ gravity, where $\phi$ represents the scalar field, and $X$ denotes its kinetic term, thus covering a vast landscape of dark energy and modified gravity models \cite{Bahamonde_2015}. In a related direction, $f(T, \phi)$ gravity, formulated from the torsion scalar $T$ and a scalar field, has been shown to admit analytical cosmological solutions in accordance with Noether symmetry requirements \cite{Duchaniya_2023,LokeshS2023}. More recently, studies of generalized $f(R, T,\phi)$ \cite{SINGH2023616} theories, where the matter sector couples through a scalar-field-dependent function of the trace of the energy momentum tensor, have revealed possible future cosmic evolutions that may culminate in a big freeze or big chill scenario.\\

The paper is organized as follows. In Section \ref{frlmgravity}, we present the mathematical formulation of the $f(R, L_m)$ gravity framework, where the action and corresponding field equations are derived, considering both linear and nonlinear dependencies on the matter Lagrangian to highlight the role of curvature--matter coupling, we introduce suitable dimensionless variables to recast the cosmological field equations into an autonomous dynamical system. In Section \ref{dsafrlm}, we performed dynamical system analysis for a detailed phase-space analysis through the identification of critical points and examination of their stability properties. In Section \ref{sec:mathematical formalism}, we extend the framework by modeling the matter Lagrangian in terms of a scalar field, thereby incorporating scalar field dynamics into the $f(R, L_m)$ gravity scenario. Section \ref{secV} is devoted to the dynamical system analysis of the generalized model in the presence of the scalar field with a self-interacting potential, where a higher-dimensional autonomous system is constructed and analyzed using appropriate techniques. Further, we discuss the cosmological implications of the model by examining the evolution of key cosmological parameters, such as the density parameters, deceleration parameter, and effective equation of state, demonstrating the existence of different evolutionary phases including matter domination and late-time accelerated expansion. Finally, in Section \ref{secvi} the main results and conclusions of the study are summarized.

\section{The  \texorpdfstring{$f(R,\mathcal{L}_m)$}{} gravity} \label{frlmgravity}
The action of $f(R,\mathcal{L}_m)$ gravity \cite{Harko_2010b} is defined as,
\begin{equation}\label{cfrl1}
S=\int f(R,\mathcal{L}_m) \sqrt{-g} d^4x ,
\end{equation}
where $\sqrt{-g}$ is the determinant of the metric tensor $g_{\mu\nu}$, $f(R,\mathcal{L}_m)$ is an arbitrary function of Ricci scalar $R$ and matter Lagrangian $\mathcal{L}_m$ respectively and $8\pi G=1$. The energy-momentum tensor is defined from \cite{LandauLifshitz1975} as,
\begin{equation}\label{stress energy momentum tensor}
T_{\mu \nu }=-\frac{2}{\sqrt{-g}}\frac{\delta \left( \sqrt{-g}\mathcal{L}_{m}\right)}{
\delta g^{\mu \nu }}\,.
\end{equation}

The value of Ricci scalar can be obtained by contracting Ricci tensor as,

\begin{align}\label{Ricci_scalar}
R=g^{\mu \nu} R_{\mu \nu}.
\end{align}

For a positive metric signature, the value of Ricci tensor is defined by using,
\begin{equation}
    R_{\mu \nu}=\frac{\partial \Gamma^{\beta}_{\mu \nu}}{\partial x^{\beta}}- \frac{\partial \Gamma^{\beta} _{\mu \beta}}{\partial x^\nu}+\Gamma^{\alpha}_{\mu \nu} \Gamma^{\beta}_{\alpha \beta}-\Gamma^{\alpha}_{\mu \beta} \Gamma^{\beta}_{\nu \alpha}.
\end{equation}    
Applying the variational principle in Eq. \eqref{cfrl1}, one can obtain the field equations of $f(R,\mathcal{L}_m)$ gravity as,

\begin{equation}\label{Eq:field1}
R_{\mu\nu}f_R+(g_{\mu\nu}\nabla_\mu\nabla^\mu-\nabla_\mu\nabla_\nu )f_R-\frac{f}{2}g_{\mu\nu}
    =\frac{1}{2}f_{\mathcal{L}_m}(T_{\mu\nu}-\mathcal{L}_mg_{\mu\nu}).
\end{equation}
 where $f\equiv f(R,\mathcal{L}_m)$, $f_R\equiv\partial f(R,\mathcal{L}_m)/\partial R$ and $f_{\mathcal{L}_m}\equiv\partial f(R,\mathcal{L}_m)/\partial\mathcal{L}_m$. We consider, $f(R,\mathcal{L}_{m})=f_{1}(R)+f_{2}(R)G(\mathcal{L}_{m})$, where $f_{1}(R)$ and $f_{2}(R)$ are arbitrary function of Ricci scalar and $G(\mathcal{L}_{m})$ is a function of matter Lagrangian density. When $f_{1}(R)=1$, $f_{2}(R)=1$ and $G(\mathcal{L}_{m})=\mathcal{L}_{m}$, Eq. \eqref{Eq:field1} reduces to the field equations of general relativity whereas for $f_{2}(R)=1$ and $G(\mathcal{L}_{m})=\mathcal{L}_{m}$ it reduces to that of $f(R)$ gravity. Also, for linear coupling of matter and geometry, $G(\mathcal{L}_{m})=1+\lambda \mathcal{L}_{m}$, where $\lambda$ is a constant. Now, the contracting form of Eq. \eqref{Eq:field1} can be written as,
\begin{equation}\label{Eq:field2}
f_{R}R+3\nabla_\mu\nabla^\mu f_{R}-2f=f_{\mathcal{L}_{m}}\left(\frac{1}{2}T-2\mathcal{L}_{m}\right).
\end{equation}
From Eq. \eqref{Eq:field1} and Eq. \eqref{Eq:field2}, one can eliminate the term $\nabla_\mu\nabla^\mu f_{R}$ and the resulting equation becomes,
\begin{equation}
f_{R}\left(R_{\mu \nu}-\frac{1}{3}R g_{\mu \nu}\right)+\frac{f}{6}-\nabla_{\mu}\nabla_{\nu}f_{R}
=\frac{1}{2}f_{\mathcal{L}_{m}}\left(T_{\mu \nu}-\frac{1}{3}(T-\mathcal{L}_{m}) g_{\mu \nu} \right).
\end{equation}
Taking the covariant divergence of Eq. \eqref{Eq:field1}, with the use of the following mathematical identity
\begin{equation}
\nabla^{\mu}\left[R_{\mu \nu }f_{R}+( g_{\mu\nu}\nabla_\mu\nabla^\mu-\nabla_{\mu}\nabla_{\nu})f_{R}-\frac{f}{2}g_{\mu \nu}\right]=0,   \end{equation}
one can obtain the divergence of energy-momentum tensor $T_{\mu \nu}$ as,
\begin{eqnarray}
\nabla^{\mu}T_{\mu \nu}&=&\nabla^{\mu}\ln[f_{\mathcal{L}_{m}}](\mathcal{L}_{m}g_{\mu \nu}-T_{\mu \nu}) \nonumber \\
&=&2\nabla^{\mu}\ln[f_{\mathcal{L}_{m}}]\frac{\partial \mathcal{L}_{m}}{\partial g^{\mu \nu}}.
\end{eqnarray}
The requirement of the conservation of the energy-momentum tensor of matter ($\nabla^{\mu}T_{\mu \nu}=0$) provides an effective functional relation between the matter and Lagrangian density. Moreover, the conservation of the energy-momentum tensor yields

\begin{equation}\label{cfrl3}
	\nabla^\mu\ln [f_{\mathcal{L}_{m}}]=0.
\end{equation}

We consider a flat Friedmann-Lema\^itre-Robertson-Walker (FLRW) metric, which assumes a homogeneous and isotropic Universe with zero spatial curvature ($\kappa = 0$). The line element in comoving spherical coordinates becomes

\begin{equation} \label{flrw}
ds^2=-dt^2+a^{2}(t)\left(dr^{2}+r^2 d\theta^2+r^2\sin^2\theta d\phi^2\right),
\end{equation} 
where $a(t)$ is the scaling factor with cosmic time $t$. The stress energy-momentum tensor for a perfect fluid can be written as,
\begin{align}\label{stressenergy}
T_{\mu \nu}=(\textit{p} + \rho ) u_{ \mu } u_{ \nu }+ \textit{p} g_{ \mu \nu },
\end{align}
where $u_{\mu}$ is the four velocity vectors along the time directions, $\rho $ is the matter energy density and $\textit{p} $ is the isotropic pressure; $u_{\mu}$ satisfy the relations $u_{\mu} u_{\nu} g^{\mu \nu }=-1$. Using Eq. \eqref{flrw} and Eq. \eqref{stressenergy} in Eq. \eqref{Eq:field1}, the field equations of $f(R, \mathcal{L}_m)$ gravity can be obtained as,
\begin{eqnarray}
3H^2f_R+\frac{1}{2}(f-f_R R-f_{\mathcal{L}_m}\mathcal{L}_{m})+3H\dot{f_R}&=&\frac{1}{2}f_{\mathcal{L}_m}\rho, \label{fieldequation1111}\\ 
\dot{H}f_R+3H^2f_R-\ddot{f_R}-3H\dot{f_R}+\frac{1}{2}(f_{\mathcal{L}_m}\mathcal{L}_{m}-f)&=&\frac{1}{2}f_{\mathcal{L}_m}p. \label{fieldequation2}
\end{eqnarray}
 where the Hubble parameter, $H=\frac{\dot{a}}{a}$, an over dot represents the ordinary derivative in time. The energy conservation for $f(R,\mathcal{L}_m)$ gravity can be written as,
\begin{equation}
\dot\rho \left(1+ \frac{2 \rho~f_{\mathcal{L}_m~\mathcal{L}_m}}{f_{\mathcal{L}_m}} \right)+3H(\rho+p)=0. \label{energy conservation}
\end{equation}
To proceed, we consider the functional form of $f(R,\mathcal{L}_m)$ as 
\begin{equation}
f(R,\mathcal{L}_m)= \frac{R}{2}+\mathcal{L}_m+\alpha \mathcal{L}_m e^\frac{\mathcal{L}_m}{\mathcal{L}_{m0}} ,   \label{form}
\end{equation}
where $\alpha$ denotes the dimensionless coupling parameter and $\mathcal{L}_{m0}$ denotes the present value of the matter Lagrangian. The geometric sector is retained in the standard Einstein-Hilbert form $\frac{R}{2}$, so that any deviation from General Relativity arises solely from the modified matter sector \cite{Harko_2010,Bhagat_2026_adf2,doi:10.1142/S0218271824500354}. The adopted matter Lagrangian is motivated by the Gaussian-process reconstruction of $g(\mathcal{L}_m)$ presented in \cite{devi2025late}. The reconstruction provides the observational basis for the adopted matter sector. Inspired by the reconstructed behaviour, we introduce an exponential correction to the linear matter contribution to account for possible nonlinear effects while preserving mathematical simplicity. The parameter $\alpha$ controls the strength of the nonlinear correction to the matter sector, the model naturally reduces to the standard matter sector in the limit $\alpha\rightarrow0$, thereby recovering the conventional Einstein-Hilbert framework in standard GR. Upon substituting the functional form in Eq.~\eqref{fieldequation1111} and Eq.~\eqref{fieldequation2} the Friedmann equations becomes
\begin{eqnarray}
    &3H^2&= \rho_m+ \alpha e^{-\rho_m/\rho_{m0}}\left(1+2\frac{\rho_m}{\rho_{m0}} \right), \label{friedmanneqnfr1}\\
    &-3H^2&-2\dot{H}=\alpha\frac{\rho_{m}^2} {\rho_{m0}}e^{-\rho_m/\rho_{m0}} \label{friedmanneqnfr2}.
\end{eqnarray}

To investigate the evolutionary behavior of the Universe, we employ the dynamical system approach. For this purpose, we introduce suitable dimensionless variables using Eq. \eqref{friedmanneqnfr1} and Eq. \eqref{friedmanneqnfr2} within the framework of $f(R,\mathcal{L}_m)$ gravity. The autonomous system of equations is obtained by differentiating these variables with respect to the e-folding number $N=\ln a$, which facilitates the determination of the corresponding critical points. We define the dimensionless dynamical variables as
\begin{equation}\label{Eq.32}
x=\frac{\rho_m}{3H^2},\ \ \ y=\frac{\rho_m}{\rho_{m0}},
\end{equation} 
where the variable $x$ represents the normalized matter density parameter, measuring the fraction of the total effective energy density associated with matter. The variable $y$ describes the matter density relative to its present value and therefore tracks its evolution throughout the cosmic history. These variables transform the cosmological equations into a dimensionless autonomous system, enabling a systematic phase-space analysis of the model. Furthermore, the chosen variables allow the effects of the exponential modification in the matter Lagrangian to be incorporated in a compact form, facilitating the identification of equilibrium solutions and their stability properties. In terms of these variables, the generalized Friedmann Eqs. \eqref{fieldequation1111}--\eqref{fieldequation2} can be rewritten as,
\begin{equation}\label{Eq:17}
1 = x + 2 \alpha~ x~ y ~e^y+\alpha~x ~e^y.
\end{equation}
Here, we eliminate the variable $e^y$ to make our system in the form of two independent variables. Now, from second Friedmann equation, we can write
\begin{equation}\label{Eq:18}
-\frac{2\dot{H}}{3H^2}=1-\alpha~ x~ y ~e^y.
\end{equation}

The autonomous system of differential equations can be obtained as, 
\begin{align}\label{Eq:20}
x' &=\frac{3 (x-1) x y (y+1) (x y+y+1)}{(2 y+1) (y ((x-1) y+x-3)-1)},\nonumber\\
y'&=\frac{3 y (x y+y+1)}{y ((x-1) y+x-3)-1},
\end{align}
where prime denotes differentiation with respect to e-folding number $N$. Eq. \eqref{Eq:20} constitute a two-dimensional autonomous dynamical system that completely determines the cosmological evolution of the model in the $(x,y)$ phase space. The evolution of $x$ describes how the matter contribution changes relative to the total effective cosmic energy, while the evolution of $y$ governs the variation of the matter density with respect to its present value. The nonlinear structure of the autonomous system originates from the exponential correction introduced in the matter sector and reflects deviations from the standard cosmological evolution.

\section{Dynamical system analysis in $f(R,\mathcal{L}_m)$ gravity} \label{dsafrlm}

The critical points can be obtained by  solving $ x' = 0 $ and $ y' = 0 $. These conditions identify the equilibrium configurations of the system corresponding to distinct cosmological phases. By examining the stability of these points, one can determine whether the associated solution represents an early-time state, a transient evolutionary phase, or a late-time attractor describing the asymptotic behaviour of the Universe. We obtain two classes of critical configurations as in TABLE--\ref{TABLE-AI}.

\begin{table}[h!]
\centering
\begin{tabular}{|c|c|c|c|c|c|}
\hline
\textbf{Critical Set} & \textbf{Equilibrium Set} & \textbf{Eigenvalues} &\textbf{Stability Condition}& \text{q} & \text{$\omega$} \\
\hline

${C}_1$ 
& $(x,0)$ 
& $\lambda_1 = 0$, \quad $\lambda_2 = -3$ & everywhere
& $\frac{1}{2}$ \quad & 0  \\

\hline

${C}_2$ 
& $\left(x,-\dfrac{1}{1+x}\right)$ 
& $\lambda_1 = 0$ \quad $\lambda_2 = \frac{3}{(-1 + x)}$ &$x\ne 1 \quad x<0$
& $-1$ & $-1$\\

\hline

\end{tabular} \label{I}
\label{Table-I}
\caption{Critical set, Stability condition, Deceleration parameter, EoS parameter.}
 \label{TABLE-AI}
\end{table} 

\noindent\textbf{Critical Set $C_1$}: The configuration ${C}_1$ represents a continuous set of equilibrium points in the phase space, given by $(x,0)$. Since this set is not isolated, the Jacobian matrix evaluated along the curve necessarily possesses at least one vanishing eigenvalue. In particular, the corresponding eigenvalues are $\lambda_1 = 0$ and $\lambda_2 = -3$, valid throughout the branch. The presence of a zero eigenvalue indicates a center direction tangent to the curve, whereas the negative eigenvalue ensures stability in the transverse direction. Consequently, the linearization theorem is insufficient to fully determine the stability properties of ${C}_1$. Therefore, the dynamical behavior in the vicinity of this critical branch must be examined using higher-order perturbative techniques. From a cosmological perspective, this branch corresponds to a matter-dominated phase, as indicated by the deceleration parameter $q = \frac{1}{2}$ and the equation of state parameter $\omega = 0$.\\

\noindent \textbf{Critical Set $C_2$}: The second equilibrium configuration is also a curve 
\begin{equation}
C_2 : (x_c,y_c)=\left(x,\frac{-1}{1+x}\right).
\end{equation}
To characterize the late-time behavior of the Universe, we consider the limit $x \to 0$, which corresponds to $\omega \to -1$. In this regime, the contribution from matter becomes negligible ($\Omega_m \approx 0$), and the cosmic dynamics is effectively dominated by the dark energy sector, leading to an accelerated expansion consistent with a de Sitter solution. In this limit, we obtain an isolated finite critical point given by $(x,y) = (0,-1)$. However, the eigenvalues of the linearized system at this point are found to be $(0,-3)$. Hence, $C_2$ is a non-hyperbolic critical point, possessing one stable direction corresponding to the negative eigenvalue and one center direction associated with the zero eigenvalue. Consequently, the standard linear stability analysis becomes inconclusive along the center direction. To fully determine the stability properties of this point, we need to employ {\bf center manifold theory} (CMT) in the neighborhood of $C_2$.

The CMT applied in nonlinear dynamical systems when linear stability analysis is inconclusive, that is, the linearization alone cannot determine the stability due to the presence of zero eigenvalues. This approach guaranties the existence of a locally invariant manifold $W^c$ called the center manifold, which is tangent to the center subspace at the equilibrium point. The essential behavior of the full non-linear system near the equilibrium is completely determined by the reduced dynamics on this center manifold \cite{Perko2013}. To proceed, a vector field $f \in C^{r}(E)$ is considered, in which $r \geq 1$ and $E \subset \mathbb{R}^{n}$ is an open neighborhood of the origin. If $f(0)=0$ and the Jacobian matrix $Df(0)$ has $c$ eigenvalues whose real parts are zero and $s$ eigenvalues whose real parts are strictly negative with $c+s=n$. So, since the presence of eigenvalues on the imaginary axis, the equilibrium at the origin becomes non-hyperbolic. With a suitable linear transformation, the system can be rewritten as,
\begin{equation}
\begin{aligned}
\dot{x} &= Ax + F(x,y), \\
\dot{y} &= By + G(x,y),
\end{aligned}
\tag{58}
\end{equation}
where $x \in \mathbb{R}^{c}$ and $y \in \mathbb{R}^{s}$ respectively correspond to the center variables and stable variables. Here, $A$ is an $c \times c$ matrix whose eigenvalues have zero real parts, while $B$ is an $s \times s$ matrix whose eigenvalues have strictly negative real parts. The nonlinear functions $F$ and $G$ satisfy
\[
F(0,0)=0, \quad G(0,0)=0, \quad DF(0,0)=0, \quad DG(0,0)=0,
\]
so that they contain only higher-order terms. Then there exist a sufficiently small $\varepsilon > 0$ and a function $g \in C^{r}(N_{\varepsilon}(0))$, with $g(0)=0$ and $Dg(0)=0$, such that the local center manifold $W^{c}$ is characterized by
\begin{equation}
W^{c} = \{(x,y) \in \mathbb{R}^{c} \times \mathbb{R}^{s} 
\; : \; y = g(x), \ |x| < \varepsilon \}.
\end{equation}
The function $g(x)$ satisfies the invariance condition
\begin{equation}
Dg(x)\big[Ax + F(x,g(x))\big] - Bg(x) - G(x,g(x)) = 0,
\tag{59}
\end{equation}
for all $|x| < \varepsilon$. 
This relation guaranties that trajectories initiated on the center manifold remain confined to it. Consequently, the dynamics restricted to the center manifold are governed by the reduced system
\begin{equation}
\dot{x} = Ax + F(x,g(x)),
\tag{60}
\end{equation}
for all $x \in \mathbb{R}^{c}$ with $|x| < \varepsilon$. The qualitative behavior and stability of the original 
$n$-dimensional system near the equilibrium are therefore  completely determined by this reduced $c$-dimensional system. Now, the Jacobian matrix of the autonomous equation given by Eq.~\eqref{Eq:20} about a critical point $(0,-1)$ along the curve $C_2$ becomes
\begin{equation}\label{jacobianpf }
\centering
J(C_2)=\left(
\begin{array}{cc}
 0 & -6 \\
 0 & -3 \\
\end{array}
\right)
\end{equation}

\noindent The corresponding eigenvalues are $\lambda_1 = 0,  \lambda_2 = -3$ and since one eigenvalue vanishes, the point is non-hyperbolic. To investigate its stability, we translate the critical point to the origin by introducing the new variables $u = x,  v = y+1$. Expanding the autonomous system around $(u,v)=(0,0)$ and keeping terms up to the lowest nonlinear order, the system can be written in the general form as
\begin{align}
u' &= f(u,v), \\
v' &= -3v + g(u,v),
\end{align}
where $f$ and $g$ contain only nonlinear terms of order $\mathcal{O}(2)$ and higher. The eigenvector associated with $\lambda_1=0$, $[1,0]^T$ defines the center subspace along the $u$-direction, while the eigenvector corresponding to $\lambda_2=-3$,$[2,1]^T$ defines a stable subspace along the $v$-direction. We assume the center manifold can be expressed locally as
\begin{equation}
v = h(u),
\end{equation}
where $h(0)=0,  h'(0)=0$. Expanding in Taylor series,
\begin{equation}
h(u)=a u^2 + b u^3 + \mathcal{O}(u^4),
\end{equation}
and imposing the invariance condition
\begin{equation}
h'(u)u' = -3h(u) + g(u,h(u)),
\end{equation}
we determine the coefficients by comparing powers of $u$. Carrying out this procedure yields $a=0$, restricting the system to the center manifold to leading order yields
\begin{equation}
u' = -3u^3.
\end{equation}
Since $u>0 \Rightarrow u'<0$ and $u<0 \Rightarrow u'>0$, the critical point $C_2=(0,-1)$ is locally asymptotically stable. We observed that after translating a generic equilibrium configuration to the origin, the system decomposes into a center direction associated with the zero eigenvalue and a transverse direction governed by the non-zero negative eigenvalue.The sign of the non-zero eigenvalue determines whether the equilibrium curve behaves as a local attractor or repeller.

\begin{figure}[H]
    \centering
    \begin{subfigure}[b]{0.45\textwidth}
        \includegraphics[width=\textwidth]{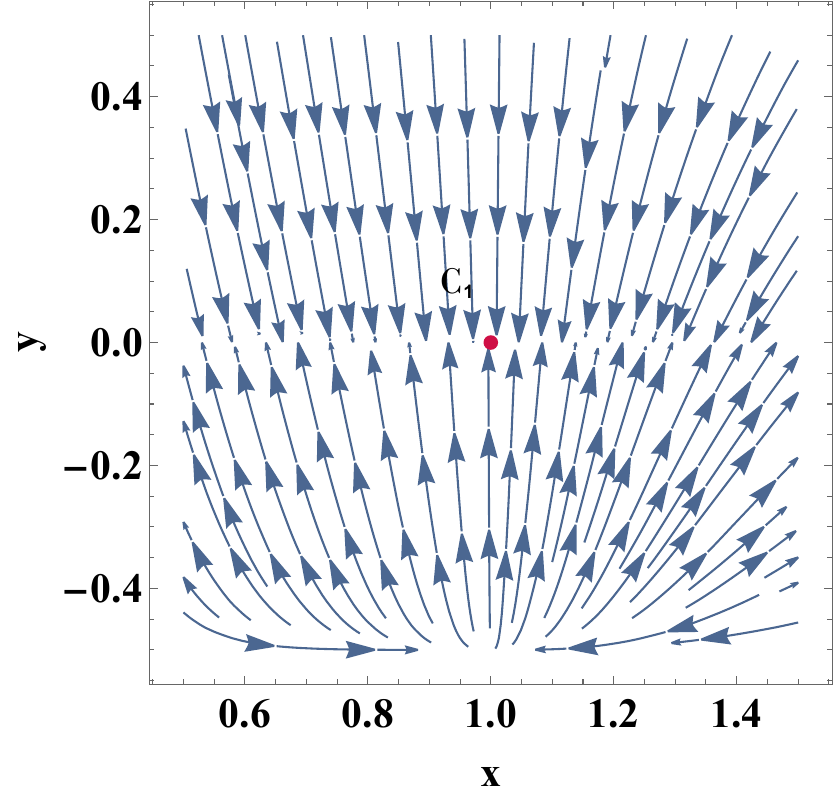}

        \label{fig:img1}
    \end{subfigure}
    \hfill
    \begin{subfigure}[b]{0.45\textwidth}
        \includegraphics[width=\textwidth]{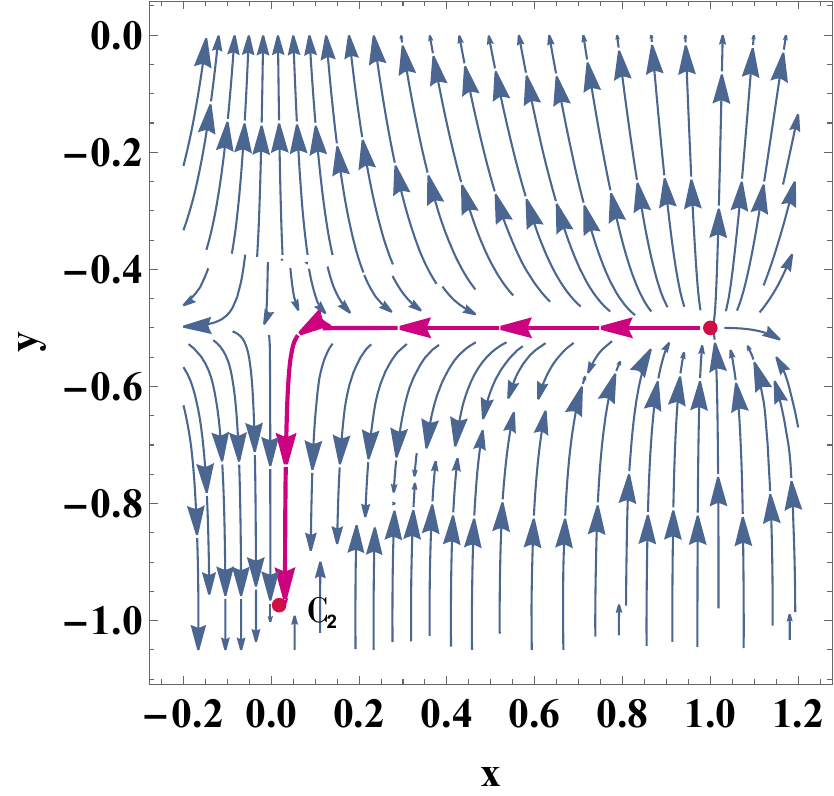}
       
        \label{fig:img2}
    \end{subfigure}
    \caption{Phase portrait for the dynamical system of the critical point $C_1${\bf(Left Panel)} and critical point $C_2${\bf(Right Panel).}}
    \label{fig:sidebyside123}
\end{figure}

\begin{figure}[H]
    \centering
    \includegraphics[width=1\linewidth]{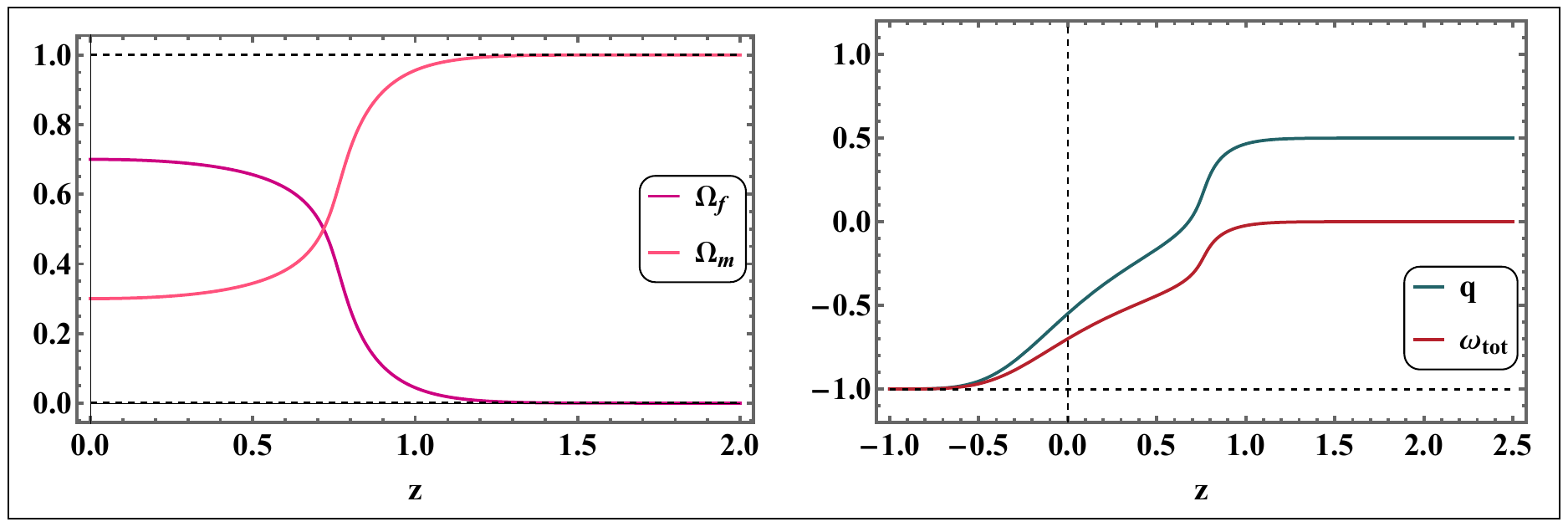}
    \caption{Evolution of density parameters ({\bf Left Panel}) and deceleration and EoS parameter ({\bf Right Panel}) in redshift.}
    \label{fig:placeholder1}
\end{figure}
Fig.~\ref{fig:sidebyside123} presents the phase space behavior of the critical configurations. The critical branch ${C}_1:(x,0)$ corresponds to a matter-dominated phase since $y=0$ along the entire curve. In this case, the nonlinear exponential contribution appearing in Eq.~\eqref{form} vanishes, and consequently the modified non-linear term effects responsible for the late-time dark-energy behaviour are absent. We examine the stability of the representative point $(1,0)$ on this branch, which corresponds to a purely matter-dominated Universe. The phase-space trajectories indicate that this point behaves as a stable configuration, implying that the cosmological evolution remains confined to the matter-dominated regime in the absence of nonlinear corrections. In contrast, the point $(1,-\frac{1}{2})$ associated with the critical branch ${C}_2:\left(x,-\frac{1}{1+x}\right)$ represents a matter-dominated state in which the nonlinear exponential contribution remains non-vanishing. The corresponding phase-space trajectories reveal saddle-type behaviour, indicating that this configuration is unstable against perturbations. The trajectories subsequently evolve toward the point $(0,-1)$, which corresponds to a de Sitter phase of accelerated expansion. Since nearby trajectories converge to this point, it acts as a stable late-time attractor of the system. Therefore, the inclusion of the nonlinear matter-sector modification significantly alters the cosmological dynamics. While the matter-dominated configuration associated with ${C}_1$ remains stable in the absence of nonlinear effects, the presence of the nonlinear term generates a dynamical pathway from a saddle matter-dominated epoch to a stable de Sitter accelerated phase, thereby providing a natural mechanism for the onset of late-time cosmic acceleration.

The cosmological nature of each critical configuration is determined by evaluating the effective equation of state parameter and the deceleration parameter along the corresponding equilibrium curves. Stable accelerating regions represent viable late-time attractor solutions, whereas unstable regions correspond to transient phases in the cosmic evolution. Fig.~\ref{fig:placeholder1} illustrates the evolution of the density parameters and the effective cosmological quantities for the exponential $f(R,\mathcal{L}_m)$ model. The {\bf Left Panel} shows that at high redshift the Universe is dominated by the matter component ($\Omega_m \approx 1$), while as the Universe expands, the effective dark energy component grows and eventually dominates at low redshift ($\Omega_f \approx 1$). The { \bf Right Panel} depicts the evolution of the total equation of state parameter and the deceleration parameter: at early times, $\omega \approx 0$ and $q>0$, corresponding to a decelerated matter-dominated phase. The deceleration parameter crosses zero at a transition redshift $z_t \approx 0.7$, indicating the onset of cosmic acceleration, with the present value $q_0 \approx -0.65$. At late times, $\omega \to -1$ and $q<0$, demonstrating that the model asymptotically approaches a de Sitter accelerating phase.

\section{The  \texorpdfstring{$f(R,\mathcal{L}_m)$}{} gravity with scalar field} \label{sec:mathematical formalism}

The action of $f(R,\mathcal{L}_m)$ gravity with minimally coupled generalized scalar field with kinetic term is \cite{Harko_2013},

\begin{equation}\label{action}
S=\int f\left(R,\mathcal{L}_m,\phi,g^{\mu\nu}\nabla_{\mu}\phi\nabla_{\nu}\phi\right) \sqrt{-g}\;d^{4}x~,
\end{equation}
where $f\left(R,\mathcal{L}_m,\phi,g^{\mu\nu}\nabla_{\mu}\phi\nabla_{\nu}\phi\right)$ is an arbitrary function of Ricci scalar $R$, matter Lagrangian density $\mathcal{L}_{m}$, scalar field $\phi$ and kinetic energy of scalar field. Applying variational principle in \eqref{action}, the field equations of the $f\left( R,\mathcal{L}_{m},\phi,(\nabla\phi)^2\right) $ gravity can be obtained as
\begin{eqnarray}\label{fieldscalar}
&&f_{R} R_{\mu \nu }+\left( g_{\mu \nu }\nabla _{\lambda }\nabla^{\lambda } -\nabla
_{\mu }\nabla _{\nu }\right) f_{R} -
\frac{1}{2}\left[
f -f_{\mathcal{L}_{m}}\mathcal{L}_{m}\right] g_{\mu \nu }=\frac{1}{2}%
f_{\mathcal{L}_{m}} T_{\mu \nu }-f_{(\nabla\phi)^2}\nabla_{\mu}\phi\nabla_{\nu}\phi
\end{eqnarray}
For brevity we denote $f_{(\nabla\phi)^2}\equiv\frac{\partial f(R,\mathcal{L}_m)}{\partial(\nabla\phi)^2}$ and consider $\mathcal{L}_m=\rho$. Varying the action with respect to the scalar field $\phi$, the corresponding Klein-Gorden equation is obtained as,
\begin{equation}
\square_{(\nabla \phi)^2}\phi=\frac{1}{2\sqrt{-g}}\frac{\partial}{\partial x^\mu}\left[f_{(\nabla \phi)^2}\sqrt{-g}g^{\mu\nu}\frac{\partial f_{\phi}}{\partial x^\nu}\right]\label{scalarfield evolution},
\end{equation}
where $\square_{(\nabla\phi)^2}$ is the generalized D'Alembert operator of $f\left(R,\mathcal{L}_m,\phi,g^{\mu\nu}\nabla_{\mu}\phi\nabla_{\nu}\phi\right)$ gravity. For a flat FLRW space-time, using \eqref{fieldscalar}, the Friedmann equation can be written as

\begin{eqnarray}
3H^2f_R+\frac{1}{2}(f-f_RR-f_{\mathcal{L}_m}\mathcal{L}_m)+3H\dot{f_R}&=&\frac{1}{2}f_{\mathcal{L}_m}\rho-f_{(-\dot{\phi^2})}\dot{\phi^2}, \label{scalarfriedmann1}\\
\dot{H}f_R+3H^2f_R-\ddot{f_R}-3H\dot{f_R}+\frac{1}{2}(f_{\mathcal{L}_m}\mathcal{L}_m-f)&=&\frac{f_{\mathcal{L}_m}p}{2}.
\label{scalarfriedmann2}
\end{eqnarray} 

The total Lagrangian of the matter-scalar gravitational field theory with the kinetic and self-interacting potential term $V(\phi)$ to be from \cite{Harko_2013} with an additional exponential term of the matter Lagrangian reconstructed from \cite{devi2025late} as,
\begin{equation}
    f(R,\mathcal{L}_m,\phi,-\dot{\phi^2})=\frac{R}{2}+\mathcal{L}_m+\alpha \mathcal{L}_m\exp^{\frac{\mathcal{L}_m}{\mathcal{L}_{m0}}}-\frac{\dot{\phi}^2}{2}+V(\phi) .\label{scalarfieldmodel}
\end{equation}
Substituting Eq.~\eqref{scalarfieldmodel} in the Friedmann Eqs. \eqref{scalarfriedmann1} and \eqref{scalarfriedmann2}, considering $\mathcal{L}_m=\rho_m$ we obtain the corresponding field equations as,
\begin{eqnarray}
    3H^2=\rho_m+\frac{3\alpha \rho_m^2}{\rho_{m0}}+\frac{\dot{\phi}^2}{2}+V(\phi)\label{linearfield1}\\
    -2\dot{H}-3H^2=p_{m}+\frac{\dot{\phi}^2}{2}-V(\phi)-\frac{\alpha \rho_m^2}{\rho_{m0}} \label{linearfield2}
\end{eqnarray}

From Eq. \eqref{linearfield1} and Eq. \eqref{linearfield2}, we obtain
\begin{eqnarray}
    \rho_{de}= \frac{3\alpha \rho_m^2}{\rho_{m0}}+\frac{\dot{\phi}^2}{2}+V(\phi)\label{rhode},\\
    p_{de}=\frac{\dot{\phi}^2}{2}-V(\phi)-\frac{\alpha \rho_m^2}{\rho_{m0}}\label{Pde}.
\end{eqnarray}

The evolution of the scalar field in Eq. \eqref{scalarfield evolution} satisfies the condition,

\begin{equation}
\ddot{\phi}+3H\dot{\phi}+\frac{dV(\phi)}{d\phi}=0.
\end{equation}
In this framework, the usual Hubble function term is modified by an additional contribution, which can act as either an effective damping or anti-damping term. This modification arises due to the non-minimal coupling between the geometric sector and the matter Lagrangian.\\

To investigate the cosmological dynamics in the framework of $f(R, \mathcal{L}_m)$ gravity with a minimally coupled generalized scalar field with a self-interacting potential term, we recast the field equations into an autonomous dynamical system by introducing appropriately dimensionless variables. We consider a scalar field $\phi$ with a self-interacting potential $V(\phi)$ of the exponential form $V(\phi)=V_0e^{-n\phi}$, and define the following Hubble-normalized variables.

\begin{equation}
x = \frac{\rho_m}{3H^2}, \quad y = \frac{\rho_m}{\rho_{m0}}, \quad \gamma=\frac{\dot{\phi}}{\sqrt{6}H},\quad \delta=\frac{\sqrt{V}}{\sqrt{3}H},\quad \lambda = -\frac{V_\phi}{V},\quad \Gamma=\frac{V_{\phi \phi}V}{{V'}^2}.\label{dynamical variables}  
\end{equation}
Here the variable $x$ represents the normalized matter density parameter and $y$ measures the evolution of the matter density relative to its present value, $\gamma$ and $\delta$ describe the kinetic and potential contributions of the scalar field, respectively, normalized by the Hubble parameter, $\lambda$ and $\Gamma$ are introduced to characterize the form of the scalar field potential. These variables transform the cosmological field equations into a dimensionless form and enable the entire cosmological dynamics to be represented as trajectories in a finite phase space. Substituting these variables in Eq. \eqref{linearfield1}, we obtain the constrained equation as,
\begin{equation}\label{123}
1=x+2\alpha xye^y+\alpha xe^y+\frac{\delta^2}{2}+\frac{\gamma^2}{2}.
\end{equation}

These dimensionless variables enable us to recast the modified Friedmann and Klein--Gordon equations into a closed autonomous system. By eliminating the exponential term $e^{y}$ using Eq.~\eqref{123}, the system reduces to a set of first-order differential equations. Consequently, the evolution equations of the dynamical system can be written as,
\begin{align}\label{scalarfieldautonomouseqn}
x' &=\frac{3 (x-1) x y (y+1) (x y+y+1)}{(2 y+1) (y ((x-1) y+x-3)-1)},\nonumber\\
y'&=\frac{3 y (x y+y+1)}{y ((x-1) y+x-3)-1},\nonumber\\
\gamma'&=-3\gamma+\sqrt{\frac{3}{2}}n\delta^2+\frac{3}{2}\gamma\left(-\frac{2\dot{H}}{3H^2}\right),\nonumber\\
\delta'&=-n\sqrt{\frac{3}{2}}\delta\gamma+\frac{3}{2}\delta\left(-\frac{2\dot{H}}{3H^2}\right).
\end{align}
The evolution of the Hubble parameter is obtained from the modified Friedmann equation~\eqref{linearfield2} as,
\begin{equation}
-\frac{2\dot{H}}{3H^2} = 1+\gamma^2-\delta^2-\alpha xye^y,
\end{equation}
where $\alpha$ is a model-dependent parameter arising from the coupling in $f(R, \mathcal{L}_m)$ gravity. For the exponential potential, we find $\Gamma = 1 $. The resulting autonomous system depends only on the chosen phase-space variables and not explicitly on cosmic time, making it suitable for critical-point analysis and stability investigations. The equilibrium points of the system correspond to cosmological solutions with distinct dynamical characteristics, while their stability determines the asymptotic behavior of the Universe.

\section{Dynamical System Analysis in generalized $f(R, \mathcal{L}_m)$ Gravity with self-interacting exponential potential}\label{secV}

The critical points can be obtained by equating $x'=0$, $y'=0$, $\gamma'=0$. The details of the critical points obtained along with their stability conditions and cosmological parameters are described in TABLE-- \ref{TABLE-II}. 

\begin{table}[h!]
\centering
\begin{tabular}{|c|c|c|c|c|c|}
\hline
\textbf{Critical Set} & \textbf{Critical Points} & \textbf{Eigenvalues} &\textbf{ Stability Condition}& \text{q} & \text{$\omega$} \\
\hline

${C}_1$ 
& $\left(x,0,2,2\right)$ 
& $\lambda_1 = 0$, ~ $\lambda_2 = -3$ , $\lambda_3 =\frac{1}{4} \left(-3 \sqrt{57}-3\right)$ ,  $\lambda_4 =\frac{1}{4} \left(3 \sqrt{57}-3\right)$ & $n=\sqrt\frac{{3}}{8}$
& $\frac{1}{2}$ \quad & 0  \\

\hline

${C}_2$ 
& $\left(0,-1,0,0\right)$ 
& $\lambda_1 = -3$ \quad $\lambda_2 = -3$ \quad $\lambda_3 =0$ \quad $\lambda_4 = 0$ &everywhere
& $-1$ &$ -1$\\
\hline
\end{tabular}
\caption{Critical set, Stability condition, Deceleration parameter, EoS parameter.}
\label{TABLE-II}
\end{table}

\noindent \textbf{Critical Set $C_1$}: The autonomous system admits a continuous family of critical points given by
\[
{C}_1=\left(x,0,2,2\right),
\]
where \(x\) be a free parameter. This represents a non-isolated critical point in the phase space. At this point, the deceleration parameter takes the value \(q=\frac{1}{2}\), which indicates a decelerating phase of the Universe. Furthermore, the effective equation of state parameter is found to be \(\omega=0\), confirming that the Universe is in a pressureless matter-dominated phase. The matter density parameter is given by \(\Omega_m = x\), and for \(x=1\), the Universe corresponds to a purely matter-dominated epoch. The corresponding eigenvalues of the Jacobian matrix are
\[
\lambda_1=0,\quad \lambda_2=-3,\quad \lambda_3=\frac{1}{4}\left(-3\sqrt{57}-3\right),\quad \lambda_4=\frac{1}{4}\left(3\sqrt{57}-3\right).
\]
It is evident that \(\lambda_2\) and \(\lambda_3\) are negative, while \(\lambda_4\) is positive, and \(\lambda_1=0\). The presence of both positive and negative eigenvalues indicates that the critical point possesses both stable and unstable directions, whereas the zero eigenvalue reflects the existence of a continuous set of equilibria. Therefore, \({C}_1\) is a non-hyperbolic saddle point. The existence of the positive eigenvalue ensures that trajectories diverge from the critical point along at least one direction, rendering it unstable. Hence, this critical point cannot represent a late-time attractor of the system, although trajectories may approach it temporarily along the stable directions.\\

\noindent \textbf{Critical Set $C_2$}: The autonomous system corresponding to Eq.~\eqref{scalarfieldautonomouseqn} evaluated at the critical point  
\[
C_2=\left(0,-1,0,0\right).
\]

The corresponding eigenvalues are
\begin{equation}
\lambda_1=-3, \qquad
\lambda_2=-3, \qquad
\lambda_3=0, \qquad
\lambda_4=0.
\label{EigenvaluesP}
\end{equation}
Since two eigenvalue vanishes, the critical point is non-hyperbolic and linear stability analysis alone is inconclusive. To determine the stability properties of this point, we shall apply CMT. The jacobian at $C_2$ is given by
\begin{equation}\label{jacobianC2 }
\centering
J(C_2)=\left(
\begin{array}{cccc}
 0 & 0 & 0 & 0 \\
 3 & -3 & 0 & 0 \\
 0 & 0 & -3 & 0\\
 0 & 0 & 0 & 0\\
\end{array}
\right).
\end{equation}

To proceed, we translate the critical point to the origin by introducing the new variables
\begin{equation}
u=x, \qquad v=y+1, \qquad p=\gamma, \qquad q=\delta .
\label{Transformation}
\end{equation}

Under this transformation the equilibrium point becomes
\[
(u,v,p,q)=(0,0,0,0).
\]

Expanding the autonomous system around the origin and retaining terms up to the lowest nonlinear order, the system can be written in the form
\[
\begin{pmatrix}
\dot{u} \\
\dot{v} \\
\dot{p} \\
\dot{q}
\end{pmatrix}
=
\begin{pmatrix}
-3 & 0 & 0 & 0 \\
0 & -3 & 0 & 0 \\
0 & 0 & 0 & 0 \\
0 & 0 & 0 & 0
\end{pmatrix}
\begin{pmatrix}
u \\
v \\
p \\
q
\end{pmatrix}
+
\begin{pmatrix}
{ F_1(u,v,p,q)} \\
{ F_2(u,v,p,q)} \\
{ F_3(u,v,p,q)} \\
{ F_4(u,v,p,q)}
\end{pmatrix},
\]

where the functions \(F_i(u,v,p,q)\) contain only nonlinear terms of order \(\mathcal{O}(2)\) and higher. The eigenvector corresponding to the zero eigenvalues determines the center subspace, which lies along the \(p\) and \(q\)-direction, while the remaining eigenvalues are negative and therefore span the stable subspace associated with the variables $u$ and $v$. According to CMT, there exists a locally invariant center manifold that can be expressed as
\begin{equation}
u=h_1(p,q), \qquad v=h_2(p,q),
\label{CenterManifold}
\end{equation}
where \(h_i(0)=0\) and \(h_i'(0)=0\). Expanding these functions in a Taylor series near the origin gives the following,
\begin{align}
h_1(p,q) &= a_1 p^2 + a_2 pq + a_3 q^2 + \mathcal{O}(|(p,q)|^3), \label{CM1} \\
h_2(p,q) &= b_1 p^2 + b_2 pq + b_3 q^2 + \mathcal{O}(|(p,q)|^3). \label{CM2}
\end{align}

Substituting these expressions into the system and imposing the invariance condition of the center manifold, the coefficients are determined by comparing terms of equal powers of \(p\) and \(q\). Carrying out this procedure leads to the reduced dynamics on the center manifold
\begin{equation}
p'=-\frac{3}{2}p^2+\mathcal{O}^3,\quad
q'=-\frac{3}{2}q^2+\mathcal{O}^3,
\label{ReducedDynamics}
\end{equation}
 Since \(q>0\) implies \(q'<0\) and \(q<0\) implies \(q'>0\), the solutions approach the origin as \(t\to\infty\), the same behavior holds for $p$ as we can see from Eq.\eqref{ReducedDynamics}. Therefore, perturbations along the center direction decay with time. Because the remaining eigenvalues are negative, perturbations in the corresponding directions also decay exponentially. Therefore, the critical point $P=\left(0,-1,0,0\right)$ is locally asymptotically stable.\\

\begin{figure}[H]
    \centering
    \begin{subfigure}[b]{0.45\textwidth}
        \includegraphics[width=\textwidth]{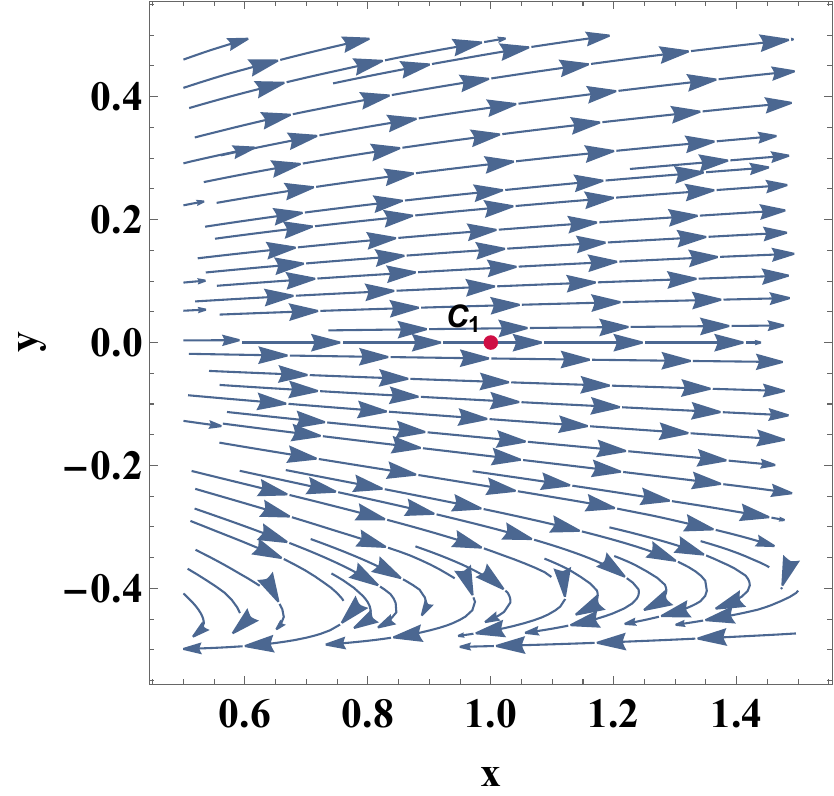}
        \label{fig:img1}
    \end{subfigure}
    \hfill
    \begin{subfigure}[b]{0.45\textwidth}
        \includegraphics[width=\textwidth]{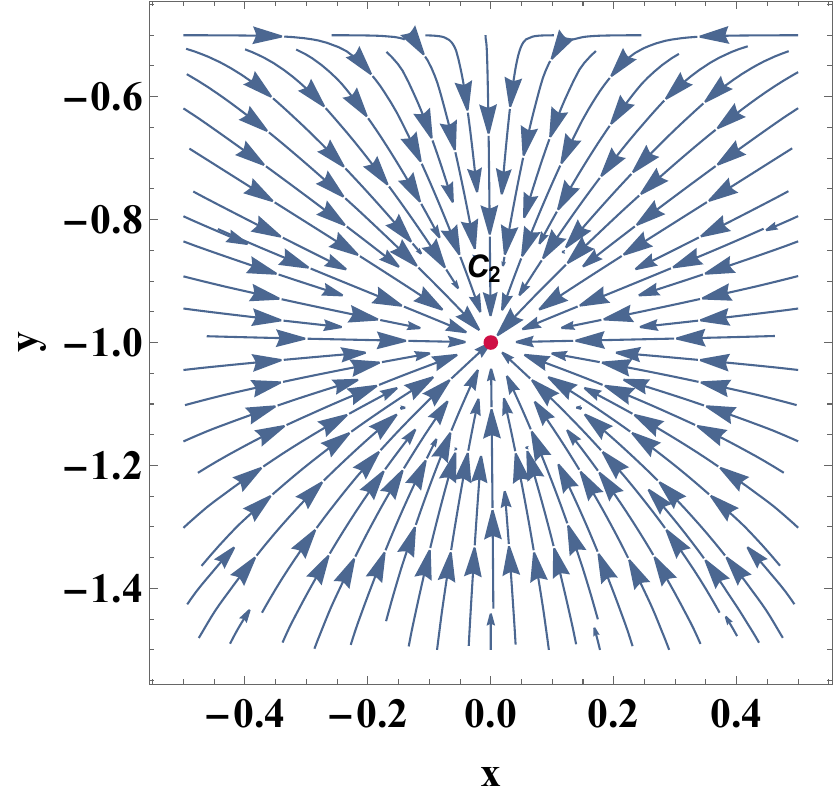}
        \label{fig:img2}
    \end{subfigure}
    \caption{ Phase portrait for the dynamical system of the critical point $C_1${\bf(Left Panel)} and critical point $C_2${\bf(Right Panel).}}
    \label{fig:sidebyside}
\end{figure}

The stability analysis of the critical points is further substantiated through the phase portrait. The critical configuration ${C}_1$, corresponding to the matter-dominated phase, exhibits saddle-like behavior due to the presence of the scalar field terms, although there is no contribution from the non-linear term, as illustrated in [Fig.~\ref{fig:sidebyside} {\bf (Left Panel)}]. The phase portrait shows that trajectories pass through this region but do not remain confined to it, indicating that the matter-dominated epoch is unstable and represents a transient stage of evolution. On the other hand, the critical point ${C}_2$, [ Fig.~\ref{fig:sidebyside} {\bf (Right Panel)}], corresponds to a de Sitter phase and displays stable attractor behavior. The trajectories in the phase space converge toward this point, confirming its stability and demonstrating that the late-time accelerated expansion of the Universe is a natural outcome of the model.

\begin{figure}[H]
    \centering
    \includegraphics[width=1\linewidth]{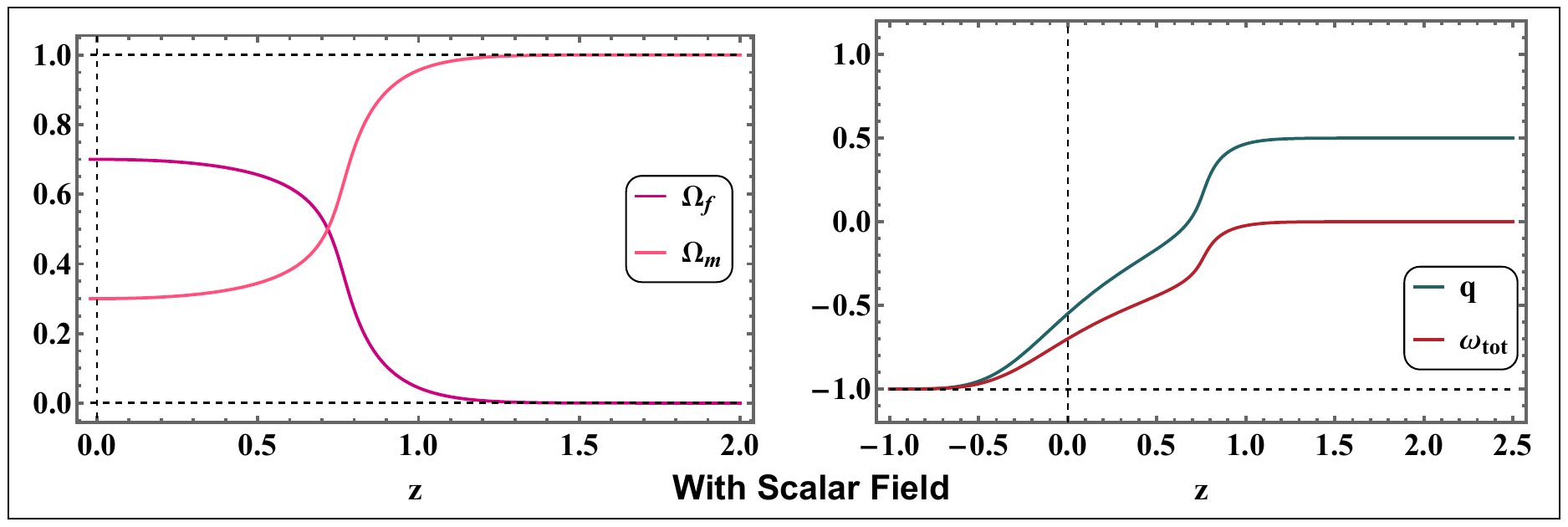}
    \caption{Evolution of density parameters ({\bf Left Panel}) and deceleration and EoS parameter ({\bf Right Panel}) in redshift.}
    \label{fig:placeholder}
\end{figure}

The autonomous system derived from the scalar field dynamics in $f(R,\mathcal{L}_m)$ gravity provides a comprehensive description of the cosmic evolution, illustrating the transition from a matter-dominated phase to a late-time accelerated epoch. This behavior is clearly depicted in Fig.~\ref{fig:placeholder} {\bf (Left Panel)}, where the evolution of the relevant cosmological parameters is shown. In particular, the effective equation of state parameter characterizes a matter-dominated phase at early times, which is further supported by the deceleration parameter taking the value $q = \frac{1}{2}$ in Fig.~\ref{fig:placeholder} {\bf (Right Panel)}. As the Universe evolves, the equation of state parameter gradually approaches $\omega \to -1$, while the deceleration parameter becomes negative and tends toward $q \to -1$, indicating the onset of accelerated expansion and the emergence of a de Sitter phase at late times. The negative value of the deceleration parameter serves as a clear signature of this accelerated behavior.

\section{Conclusion}\label{secvi}

We have presented a comprehensive investigation of cosmological dynamics within the framework of $f(R, \mathcal{L}_m)$ gravity using dynamical system techniques. By considering a specific functional form involving an exponential dependence on the matter Lagrangian, we reformulated the modified Friedmann equations into an autonomous system through the introduction of suitable dimensionless variables. This approach enabled a systematic phase-space analysis and facilitated the identification of equilibrium configurations corresponding to different cosmological epochs. In the first part of the analysis, where only the curvature--matter coupling is considered, the autonomous system admits non-isolated critical curves rather than discrete fixed points. The presence of zero eigenvalues in the Jacobian matrix renders all equilibrium configurations non-hyperbolic, making linear stability analysis insufficient. To overcome this limitation, we employ CMT, which allowed us to capture the true nonlinear behavior near the critical configurations. The analysis shows that the physically relevant critical point behaves as a locally asymptotically stable attractor. From a cosmological perspective, this corresponds to a late-time accelerating phase of the Universe. The evolution of the deceleration parameter and the effective equation of state parameter clearly demonstrate a transition from an early matter-dominated decelerating phase to a late-time accelerated expansion. In second part, we extended the model by incorporating a minimally coupled generalized scalar field with a kinetic term and a self-interacting exponential potential. This generalization significantly enriches the dynamical structure of the system. The resulting higher-dimensional autonomous system allows for a wider class of cosmological behaviors, including matter-dominated phases, scaling solutions, and accelerated attractor solutions. The critical point analysis reveals that one of the equilibrium configurations acts as a stable attractor, as confirmed through the application of CMT. This attractor corresponds to a dark energy dominated phase where the scalar field and the curvature--matter coupling jointly drive the acceleration of the Universe.\\

An important outcome of this study is the distinct role played by the nonlinear matter-sector correction and the scalar field in shaping the cosmological dynamics. In the absence of the scalar field, the matter-dominated configuration corresponding to a vanishing nonlinear exponential contribution behaves as a stable equilibrium state, causing the cosmological evolution to remain confined to the matter-dominated regime. However, upon the inclusion of the minimally coupled scalar field, the dynamical structure is significantly altered, and the matter-dominated configuration becomes a saddle-type solution. This allows the cosmological trajectories to evolve away from the matter era and approach a stable late-time accelerated attractor. Interestingly, in both scenarios, the stable accelerated solution is characterized by the critical configuration $y=\frac{\rho_m}{\rho_{m0}}=-1$, corresponding to a dark-energy-dominated de Sitter-like phase. These results demonstrate that the combined effects of the scalar field and the nonlinear matter-sector modification provide a natural mechanism for realizing the transition from matter domination to late-time accelerated expansion without the need for an explicit cosmological constant. The exponential form of the coupling function and the scalar field potential play a crucial role in determining the stability and cosmological viability of the model. The phase space trajectories further support the existence of a smooth transition from a decelerated regime to an accelerated regime, with the late-time dynamics approaching a de Sitter-like solution characterized by $\omega \to -1$ and a negative deceleration parameter. It successfully explains the sequence of cosmological phases, including the matter-dominated era and the late-time accelerated expansion. The results indicate that curvature--matter coupling and scalar field dynamics can serve as promising alternatives to the standard $\Lambda$CDM model. The present results are also qualitatively consistent with several recent studies in modified gravity that have been confronted with observational datasets \cite{SULTANA2026100620, KOUSSOUR2025117123}. In particular,\cite{Sultana2026} recent analyses of exponential-potential quintessence models and reconstructed dark energy scenarios in extended gravity frameworks have shown that viable cosmological models should exhibit a transition from a matter-dominated epoch to a late-time accelerated phase while remaining close to the $\Lambda$CDM behaviour at late times. In agreement with these findings, the stable attractor solutions obtained in the present $f(R,\mathcal{L}_m)$ model lead to an accelerated expansion phase with an effective equation of state approaching the dark-energy regime. Although the current work is based on a dynamical systems approach and does not include direct observational constraints, the obtained cosmological evolution is qualitatively compatible with the behaviour reported in recent observationally tested modified gravity models. Finally, this work opens several directions for future research. One may extend the present analysis by considering more general forms of the coupling function and scalar field potential, confronting the model with observational data, or studying perturbations and structure formation within this framework. Such investigations would further clarify the role of $f(R, \mathcal{L}_m)$ gravity in explaining the dynamics of the Universe and its compatibility with precision cosmology.

\section*{Acknowledgement} BM acknowledges the support of Council of Scientific and Industrial Research (CSIR) for the project grant (No. 03/1493/23/EMR II). RB acknowledges the financial support provided by the University Grants Commission (UGC) through Senior Research Fellowship UGC-Ref. No.: 211610028858 to carry out the research work. The authors thank IUCAA, Pune (India) for providing support in the form of an academic visit during which this work is accomplished.

\section*{References}
\bibliography{refs}
\end{document}